\pdfoutput=1

\documentclass[twocolumn,aps,prl,superscriptaddress]{revtex4-1}

\usepackage{graphicx}
\usepackage{epsfig}
\usepackage{amsfonts}
\usepackage{amsmath}
\usepackage{amssymb}
\usepackage{color}
\usepackage{multirow}
\usepackage[colorlinks=true,citecolor=blue,urlcolor=blue]{hyperref}

\begin{document}


\title{Classical Physics and the Bounds of Quantum Correlations}


\author{Diego Frustaglia}
\email{frustaglia@us.es}
\affiliation{Departamento de F\'{\i}sica Aplicada II, Universidad de Sevilla, E-41012 Sevilla, Spain}

\author{Jos\'e P. Baltan\'as}
\affiliation{Departamento de F\'{\i}sica Aplicada II, Universidad de Sevilla, E-41012 Sevilla, Spain}

\author{Mar\'{\i}a C. Vel\'azquez-Ahumada}
\affiliation{Departamento de Electr\'onica y Electromagnetismo, Universidad de Sevilla, E-41012 Sevilla, Spain}

\author{Armando Fern\'andez-Prieto}
\affiliation{Departamento de Electr\'onica y Electromagnetismo, Universidad de Sevilla, E-41012 Sevilla, Spain}

\author{Aintzane Lujambio}
\affiliation{Departamento de Electr\'onica y Electromagnetismo, Universidad de Sevilla, E-41012 Sevilla, Spain}

\author{Vicente Losada}
\affiliation{Departamento de F\'{\i}sica Aplicada I, Universidad de Sevilla, E-41011 Sevilla, Spain}

\author{Manuel J. Freire}
\affiliation{Departamento de Electr\'onica y Electromagnetismo, Universidad de Sevilla, E-41012 Sevilla, Spain}

\author{Ad\'an Cabello}
\email{adan@us.es}
\affiliation{Departamento de F\'{\i}sica Aplicada II, Universidad de Sevilla, E-41012 Sevilla, Spain}


\date{\today}


\begin{abstract}
A unifying principle explaining the numerical bounds of quantum correlations remains elusive despite the efforts devoted to identifying it. Here we show that these bounds are indeed not exclusive to quantum theory: for any abstract correlation scenario with compatible measurements, models based on classical waves produce probability distributions indistinguishable from those of quantum theory and, therefore, share the same bounds. We demonstrate this finding by implementing classical microwaves that propagate along meter-size transmission-line circuits and reproduce the probabilities of three emblematic quantum experiments. Our results show that the ``quantum" bounds would also occur in a classical universe without quanta. The implications of this observation are discussed. 
\end{abstract}

\maketitle


\emph{Introduction.---}In quantum theory (QT), an intriguing set of numbers appear in correlation experiments with compatible observables. An example is the celebrated Tsirelson bound $2 \sqrt{2}$ \cite{Tsirelson80},
the maximum quantum violation of the Clauser-Horne-Shimony-Holt (CHSH) Bell inequality \cite{CHSH69} recently ``touched'' in experiments \cite{PJCCK15}. Another one is the maximum quantum violation of the Klyachko-Can-Binicio\u{g}lu-Shumovsky noncontextuality inequality \cite{KCBS08}, $\sqrt{5}$. Popescu and Rohrlich \cite{PR94} made the first attempts to identify a principle behind all of these numbers. Recent works have found different principles enforcing $2 \sqrt{2}$ \cite{PPKSWZ09,NW09,Cabello15} and $\sqrt{5}$ \cite{Cabello13} among other quantum bounds \cite{note-1,BG03}. Still, it is an open question as to whether a single principle can grasp them all.

Remarkably, all principles able to work out quantum bounds thus far,
i.e., information causality \cite{PPKSWZ09}, macroscopic locality \cite{NW09}, and exclusivity \cite{Cabello13,Cabello15}, are also satisfied by classical physics. This means that classical physics cannot surpass the quantum bounds by turning to extra resources such as superluminal communication \cite{BCT99} and/or memory \cite{KGPLC11} without violating these principles. It also raises the question of whether classical physics can {\em saturate} the quantum bounds. A positive answer would indicate that none of these numbers are specific to QT and that they would still be natural in a fundamentally classical (nonquantum) world. 

In this Letter we show that physical models employing classical waves to produce discrete events lead to probability distributions indistinguishable from those of QT and therefore saturate all the ``quantum" bounds. We benefit from a universal mapping between correlation experiments on quantum systems and a protocol based on the detection of intensities of classical waves propagating in circuits with an appropriate configuration. The mapping is universal in the sense that it applies to any abstract correlation inequality with compatible measurements (i.e., any noncontextuality (NC) inequality \cite{Cabello08}).

We implement this mapping in a series of three experiments with classical microwaves propagating along meter-size transmission-line networks. Each of them reproduces the probability distribution of an emblematic experiment in QT. This illustrates how notions such as repeatability, nondisturbance, incompatibility, and contextuality can be defined with classical waves. It also demonstrates that the corresponding correlations share the exact bounds shown by quantum correlations.

Our results prove that the bounds of quantum correlations are not a hallmark of QT since they can \emph{all} be attained by a universal classical approach with corresponding resources (such as memory). We discuss these and other implications at the end.


\begin{figure}[htb!]
\centering
\includegraphics[scale=0.38]{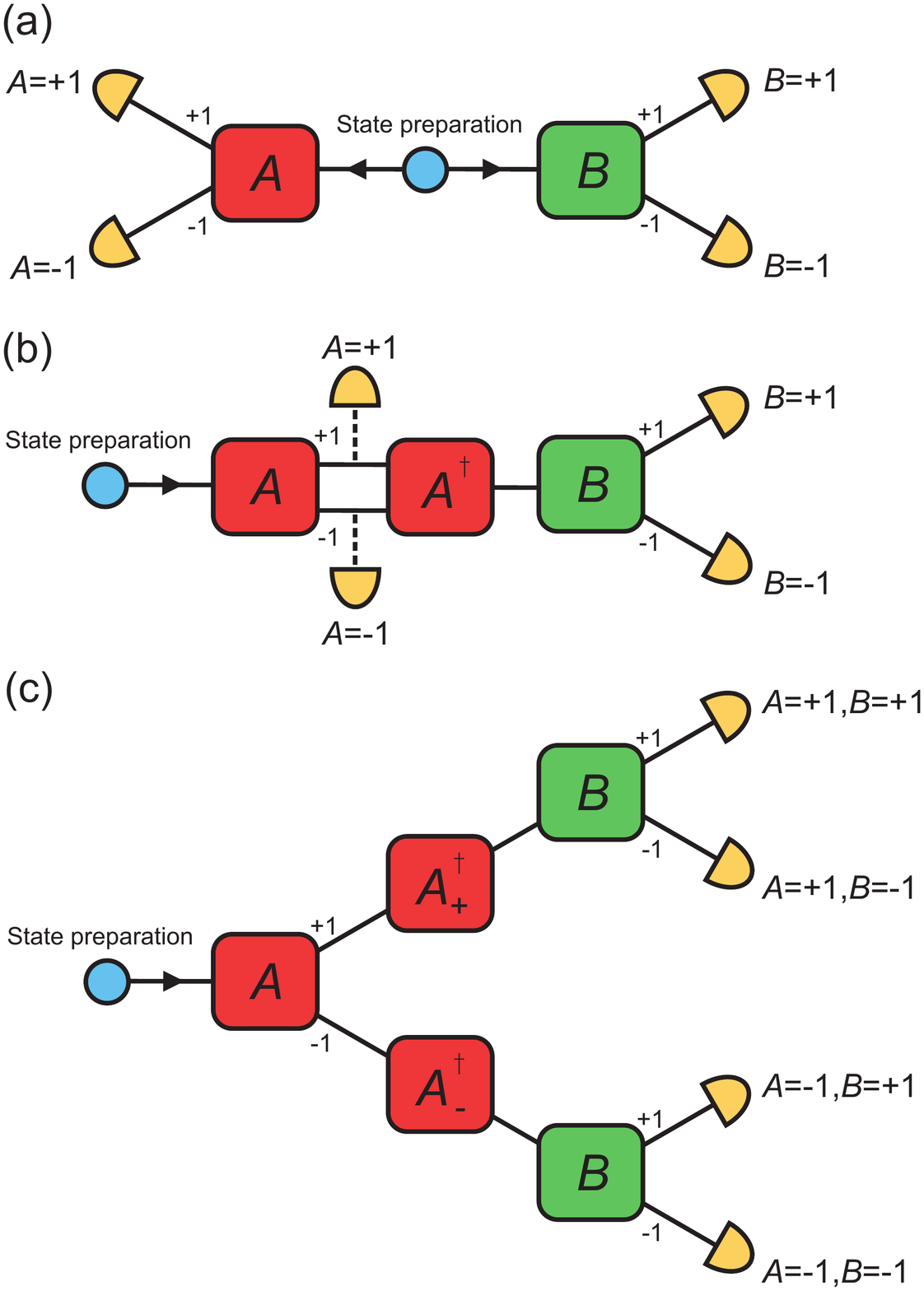}
\caption{\label{Fig1}
(a) Bell inequality experiment. A source emits a composite system and, on each subsystem, a unitary operation and a demolition measurement with two possible outcomes, $1$ or $-1$, is performed.
(b) Contextuality experiment on individual quantum systems. A source emits a single system and a sequence of compatible measurements is performed. Each measurement consists of a unitary operation $A$, followed by a nondemolition measurement that records the outcome in an external device, and a unitary operation $A^\dag$ that recomposes the input. Together, these three steps implement what in QT is represented by a self-adjoint operator: something that not only provides an outcome satisfying the Born rule but also prepares a state satisfying the L\"uders rule. The last measurement can be a demolition measurement.
(c) Tree configuration avoiding nondemolition measurements. The outcome of each intermediate measurement is encoded in an extra spatial path: upper paths for outcomes $1$ and lower paths for outcomes $-1$. In our experiments, these paths are transmission-line wires.}
\end{figure}


\emph{Assumptions and model.---}Reproducing quantum probabilities means reproducing two key features of QT: incompatibility and contextuality. Incompatibility is the impossibility of assigning a joint probability to the results of certain measurements independently of the order in which these measurements are performed. Contextuality is the impossibility of explaining joint probabilities by assuming that measurement results correspond to predefined values \cite{Bell66,KS67}. The violation of Bell inequalities is a special form of contextuality that involves causally disconnected measurements \cite{Bell64}. However, Bell-inequality experiments can be formally mapped into experiments involving sequential measurements by replacing spacelike separation with compatibility.
In this way, any Bell-inequality experiment [see Fig.~\ref{Fig1}(a)] is just a sequential contextuality experiment [see Fig.~\ref{Fig1}(b)] in a certain reference frame.

The inputs of a contextuality experiment are a physical system in a given state and a set of observables. Only compatible observables are measured in each trial. The output of the experiment is the list of joint probabilities for the compatible subsets of observables. Contextuality is manifested when these probabilities violate a NC inequality.

Our goal is to build physical models that produce any {\em distribution of probability} that can be obtained with quantum observables of a discrete spectrum measured sequentially on individual quantum systems by using classical electromagnetic waves and a suitable definition of discrete events. We adopt the standard definition of quantum observables, namely, those represented in QT by self-adjoint operators \cite{vonNeumann32}. The preparation and measurement devices are constructed by analogy with QT. For that, we remind the reader that any quantum observable can be implemented as a unitary operation $A$ followed by a nondemolition measurement and the unitary operation $A^\dag$ that recomposes the input [as in the intermediate step in Fig.~\ref{Fig1}(b)]. On an electromagnetic wave, any discrete finite-dimensional unitary operator can be realized by using a sequence of two-dimensional beam splitters \cite{RZBB94}. This allows us to construct any unitary operator needed for the preparation and measurement devices.

In sequential measurements, the intermediate outcomes are typically recorded externally [Fig.~\ref{Fig1}(b)]. We instead encode them in an extra, inner degree of freedom unfolding an arborescent network [Fig.~\ref{Fig1}(c)]. This same approach was used with success to demonstrate contextuality with path- and polarization-encoded single photons (where nondemolition measurements are not possible) \cite{ARBC09}. 

We implement the sequential arrangement sketched in Fig.~\ref{Fig1}(c) experimentally by probing classical microwave states encoded in spatial modes propagating along transmission-line
tree circuits. There, we observe that the joint probability distributions found in QT with individual systems can be reproduced with models in which events originate from microwave intensities detected at the output ports. For example, events
can be defined either as clicks of detectors that are triggered only after an energy threshold
$E_0\pm \Delta E$ is surpassed or as the outcomes after throwing a dice
(with as many faces as output ports) loaded according to the
normalized intensity distribution; see the Supplemental Material for
further details. Notice that none of these models assume or need the
existence of quanta. However, they produce probabilities
leading to the same relations of incompatibility between measurements
and to the same violations of NC inequalities found in QT. 


\emph{Tested inequalities.---}We address three emblematic experiments in QT. The first target is the maximum violation of the CHSH-Bell inequality \cite{CHSH69} (i.e., the Tsirelson bound \cite{Tsirelson80}). If we consider four observables $A,B,a$, and $b$ with two possible outcomes, $+1$ and $-1$, any model with noncontextual outcomes must satisfy $E \le 2$, with
\begin{equation}
E \equiv \langle AB \rangle +
\langle bB \rangle +
\langle Aa \rangle -
\langle ba \rangle,
\label{CHSH}
\end{equation}
where $\langle AB\rangle$ is the average of the product of the outcomes of $A$ and $B$, with $A$ and $B$ compatible. However, there are quantum states violating the CHSH inequality up to $E=2\sqrt{2} \approx 2.828$. This can be obtained with the following observables of a two-qubit system: $A=\sigma_z \otimes \mathbb{I}$, $B=\mathbb{I} \otimes \sigma_z$, $a=\mathbb{I} \otimes \sigma_x$, and $b=\sigma_x \otimes \mathbb{I}$ (where, e.g., $\sigma_z \otimes \mathbb{I}$ is the tensor product of the Pauli $z$ operator acting on the first qubit times the identity in the Hilbert space of the second qubit), and with the initial state $|\Psi_{\rm CHSH} \rangle = \left[|00\rangle-|11\rangle+(\sqrt{2}-1)(|01\rangle+|10\rangle)\right]/(2\sqrt{2-\sqrt{2}})$ [this choice connects Eq.~(\ref{CHSH}) with the following inequalities].

Interestingly, there is a gap between the quantum limit to $E$ and the maximum allowed by the no-signaling principle \cite{PR94}. Substantial effort has been put into understanding this limit \cite{PPKSWZ09,NW09,Cabello15}. This gap vanishes when we consider two more dichotomic observables, $C$ and $c$, and extend the CHSH Bell inequality into $M \le 2$ \cite{Mermin90}, with
\begin{equation}
M \equiv \langle ABc \rangle +
\langle bBC \rangle +
\langle AaC \rangle -
\langle bac \rangle.
\label{MB}
\end{equation}
In QT, the value $M=4$ can be obtained in a three-qubit system with the same observables as before, plus $C=\mathbb{I} \otimes \mathbb{I} \otimes \sigma_z$ and $c=\mathbb{I} \otimes \mathbb{I} \otimes \sigma_x$, and preparing a particular Greeberger-Horne-Zeilinger-like (GHZ-like) state $|\Psi_{\rm GHZ} \rangle$ \cite{GHZ89}.

In the two previous inequalities, once measurements are fixed, the quantum violation only occurs for certain quantum states. However, if we consider five additional dichotomic observables, $D,d,\alpha,\beta$, and $\delta$, and extend the CHSH Bell inequality into $\chi \le 4$, with
\begin{equation}
\chi \equiv \langle ABD \rangle +
\langle abd \rangle +
\langle \alpha \beta \delta \rangle +
\langle A a \alpha \rangle +
\langle B b \beta \rangle -
\langle D d \delta \rangle,
\label{PM}
\end{equation}
then there is a set of quantum observables on a two-qubit system for which $\chi=6$, no matter which quantum state the system is prepared in \cite{Cabello08}. This is called quantum state-independent contextuality and it can be observed with the same $A$, $B$, $a$, and $b$ used before, plus the observables $D=\sigma_z \otimes \sigma_z$, $d=\sigma_x \otimes \sigma_x$, $\alpha =\sigma_z \otimes \sigma_x$, $\beta =\sigma_x \otimes \sigma_z$, and $\delta=\sigma_y \otimes \sigma_y$.


\emph{Experimental setup.---}We build a series of networks of the type illustrated in Fig.~\ref{Fig1}(c) for each test. In those cases in which we need to measure sequences of two observables, as when testing Eq.~(\ref{CHSH}), the experimental setup looks exactly like Fig.~\ref{Fig1}(c). When we measure sequences of three observables, as in Eqs.~(\ref{MB}) and (\ref{PM}), the setup incorporates extra splittings.


\begin{figure}[tb]
\centering
\includegraphics[scale=0.33]{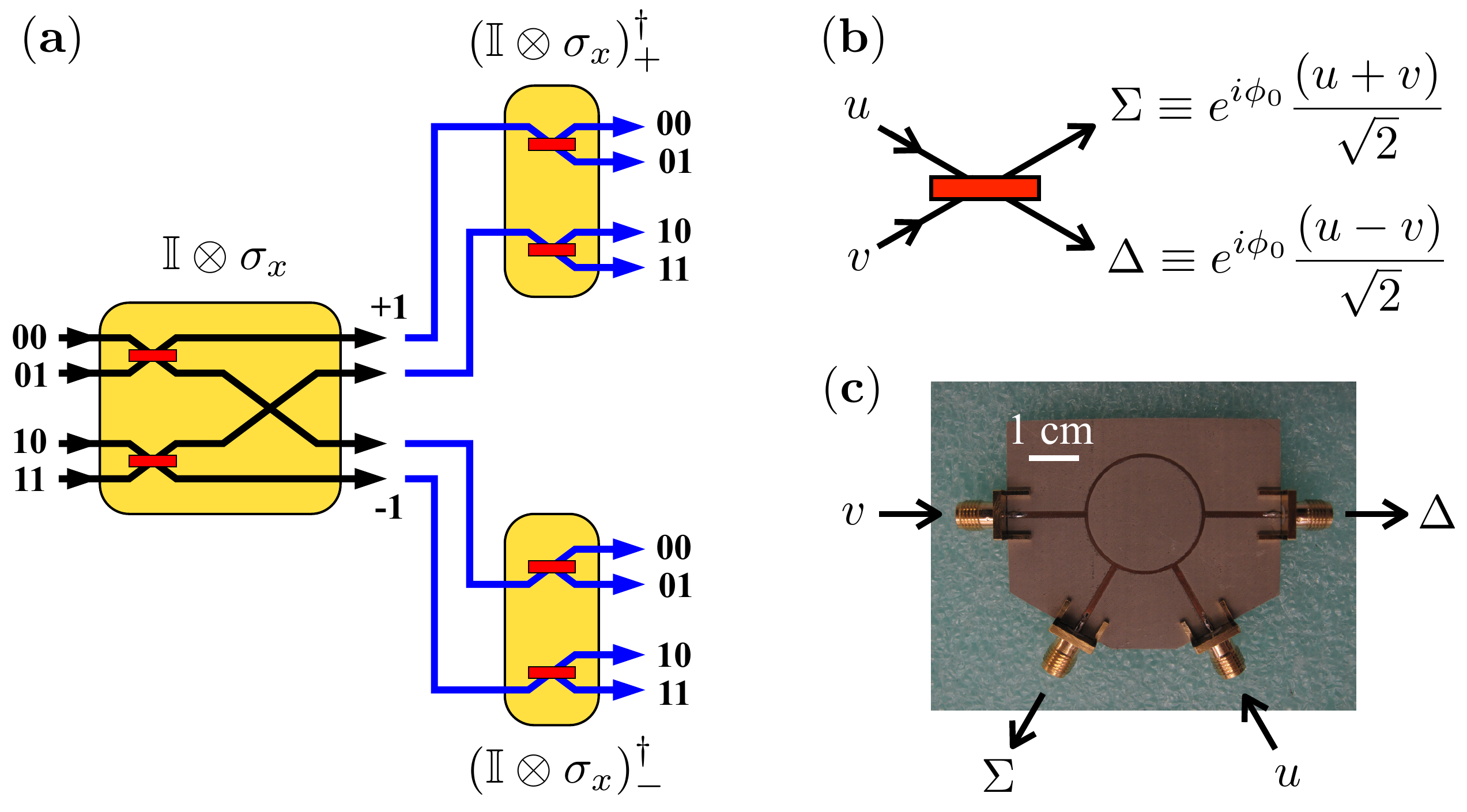}
\caption{\label{Fig2} Path encoding. (a) Schematic representation of the observable $a= \mathbb{I} \otimes \sigma_x$ used for testing inequalities (\ref{CHSH}) and (\ref{PM}). Each propagating channel identifies with an element of the Hilbert space basis. Operators $(\mathbb{I} \otimes \sigma_x)_+^\dag$ and $(\mathbb{I} \otimes \sigma_x)_-^\dag$ recompose the incoming state. (b) Beam splitter used as a building block to process incoming signal amplitudes $u$ and $v$ scattered as $\Sigma$ and $\Delta$. (c) Rat-race (hybrid ring) coupler acting as a classical microwave beam splitter in our experimental setup, designed to work at $2.45$ GHz.}
\end{figure}


The circuit implementation of the operators is built upon two basic elements: coaxial-cable segments of equal electrical length and hybrid-ring (rat-race) couplers used as beam splitters. These elements are designed to work at $2.45$ GHz (the usual frequency of consumer microwave ovens). See Figs.~\ref{Fig2}(b) and \ref{Fig2}(c) and the Supplemental Material.

We designed the devices to produce the initial states and to test the desired observables as we would do for a quantum system defined by four or eight spatial modes of a photon. We used the one-to-one correspondence between unitary transformations and beam-splitter arrangements of Ref.~\cite{RZBB94}. For example, a path representation of operator $a=\mathbb{I} \otimes \sigma_x$ is depicted in Fig.~\ref{Fig2}(a). See Figs.\ S1 and S2 in the Supplemental Material for the circuit implementation of all other operators.

The classical equivalents to quantum states are multichannel microwave signals propagating along independent waveguides with well-defined relative phases. Each classical microwave channel is identified with an element of the Hilbert space basis. The states are produced from a single microwave source by coherent splitting. For example, the classical equivalent of a singlet state $|\Psi_-\rangle = (|10\rangle-|01\rangle)/\sqrt{2}$ is created by injecting the microwave signal into the $v$ port of a hybrid-ring beam splitter [see Figs.~\ref{Fig2}(b) and \ref{Fig2}(c)] and identifying the output ports $\Sigma$ and $\Delta$ with the corresponding input channels 10 and 01 in, e.g., Fig.~\ref{Fig2}(a).

The classical analog of the state $|\Psi_{\rm CHSH} \rangle$ which reaches the Tsirelson bound is produced with the help of an unequal split branch line coupler; see Fig.\ S3 in the Supplemental Material.

The state $|\Psi_{\rm GHZ} \rangle$ is the only common eigenstate with eigenvalue $+1$ of the operators $\sigma_x \otimes \sigma_z\otimes \sigma_z$, $\sigma_z \otimes \sigma_x \otimes \sigma_z$, and $\sigma_z \otimes \sigma_z \otimes \sigma_x$. Therefore, to produce the classical analog of $|\Psi_{\rm GHZ} \rangle$ we arrange sequentially the measurement devices representing these operators and select the corresponding outcomes for an arbitrary incoming signal; see Fig.\ S4 in the Supplemental Material.

The signal was generated and measured by an automatic vector network analyzer; see Fig.\ S5 in the Supplemental Material. The experimental outcomes are the normalized transmission coefficients identified as joint probabilities distributed over the output ports. The effect of microwave power loss along the circuit is equilibrated by the symmetric design of the tree network [Fig.~\ref{Fig1}(c) and Fig.\ S6 in the Supplemental Material, for example].

We perform a series of control tests to establish that the conditions of compatibility are satisfied up to an acceptable degree. Concretely, we check the following: (i) Marginal probabilities are context independent. This is tested with different states by placing identically built circuits for each operator in every possible context. (ii) Marginal probabilities are order independent. This is tested by placing the circuits in all possible orders. (iii) Results are repeatable. This is tested by measuring sequences like $AAA$ and checking to see that the results of all of the measurements are equal. (iv) Measurements are nondisturbing. This is tested by measuring sequences like $ABA$, where $B$ is compatible with $A$, and checking to see that the results of the first and second measurements of $A$ are equal. See the Supplemental Material.

Even when all previous requirements are satisfied by design, imperfections in the sample fabrication and assemblage inevitably lead to small signal leakages into paths that should not be taken in the ideal case. This is accounted for by the corresponding deviation rates that translate into an increment of the upper bounds of inequalities (\ref{CHSH}), (\ref{MB}), and (\ref{PM}). See the Supplemental Material.


\emph{Experimental results.---}Our experiment with the classical microwave state analog to $|\Psi_{\rm CHSH} \rangle$ brought the result $E=2.78(14)$. This represents a clear-cut violation of inequality (\ref{CHSH}), which is in agreement with the results obtained in experiments with two-photon states (e.g., $E=2.697(15)$ \cite{ADR82}). Indeed, our result is closer to the quantum maximum, $E=2.828$, than those from homologous experiments testing inequality (\ref{CHSH}) on single-photon [$E=2.595(15)$ \cite{MWZ00}] and single-neutron states [$E=2.365(13)$ \cite{GDSLH14}].

In the experiment with the classical microwave GHZ-like state, we obtained $M=3.93(11)$, showing a large violation of inequality (\ref{MB}). Curiously, this experimental result is much closer to the maximum predicted by QT, $M=4$, than the results of experiments with three-photon states and spacelike separation [$M=2.77(8)$ \cite{EMFLHYPBPSRGLWJR14}] or with single-photon [$M=3.551(13)$ \cite{MWZ00}] and single-neutron states [$M=2.558(4)$ \cite{HLBDSR10}]. Moreover, recent realizations with classical light obtained a relatively small violation $M=2.62(5)$ \cite{BSCGHK15}.

We tested $\chi$ over $11$ different input states which are the classical analogs of pure quantum states with different degrees of entanglement, from separable to maximally entangled states (listed in Table S1 of the Supplemental Material). The experimental results show a clear state-independent violation of the NC inequality (\ref{PM}) with an average value $\chi=5.93(24)$, which is significantly closer to the QT prediction for an ideal experiment, $\chi=6$, than the one obtained in previous experiments with single-ion [with $\chi$ ranging from $5.23(5)$ to $5.46(4)$ \cite{KZGKGCBR09}] and single-photon states [with an average value of $\chi=5.4550(6)$ \cite{ARBC09}].


\emph{Discussion.---}Here, we have shown that, under the precise terms
defined in the mapping between Figs.~\ref{Fig1}(a) and \ref{Fig1}(b) into
Fig.~\ref{Fig1}(c), classical-wave protocols and QT produce
the same set of correlations.

Unlike \emph{ad hoc} classical models reproducing some nonlocality
\cite{BCT99} and contextuality experiments
\cite{KGPLC11,LaCour09,PL14,Blasiak15}, our approach is
universal and tight: it reproduces any possible structure
of compatibility or incompatibility and any form of contextuality that is
possible in QT, and fails to produce any nonquantum distribution. This
shows that quantum correlations can be \emph{universally} recreated with classical systems at the expense of some extra resources. In our case,
the extra memory needed to display contextuality \cite{KGPLC11}
is provided naturally by the network branches (used to store the outcomes of partial measurements) and the microwave phase taken from one generation of observables to the next one.

Our results have several implications: (I) The bounds of quantum correlations are not distinctive of QT. Hence, the principles determining the extent of quantum correlations (\emph{necessarily} shared by universal models employing classical waves plus memory) are \emph{insufficient} to grasp QT. 
This also means that even if quantum systems would not exist (and classical fields would be the fundamental physical objects), the so-called quantum bounds would still arise naturally.
(II) The characteristic trait of QT rely on the fact that the quantum bounds are achieved \emph{without} employing extra resources such as memory. Therefore, the principles needed to fully derive QT (in the spirit of Refs. \cite{Hardy01,Hardy11,DB11,MM11,CDP11}) should account for that.
(III) Our model has at disposal more memory than strictly needed to simulate the quantum bounds: it has one bit of memory for each dichotomic decision, which is more than needed to simulate quantum probabilities (with this memory, one could simulate, e.g., a nonlocal box \cite{PR94}). However, it stops right at the quantum bound, respecting the constraints imposed by information causality \cite{PPKSWZ09} and exclusivity \cite{Cabello13} (while a simulation of a nonlocal box would indicate their violation). This suggests that the bounds are not related to the availability or the amount of extra resources.
(IV) Finally, the fact that our experiments do not require any special conditions of isolation or control beyond phase coherence demonstrates that quantumlike probabilities and correlations can emerge in other classical supports with an appropriate network structure allowing the coherent propagation of wave signals. The possibilities run from artificial networks to biological ones. This also warns un about potential quantumlike features of classical origin that could be wrongly taken as actual quantum effects in complex systems. 


This work was supported by Project No.\ FIS2011-29400 and No. FIS2014-60843-P (MINECO, Spain) with FEDER funds and the FQXi large grant project ``The Nature of Information in Sequential Quantum Measurements.'' We thank E.\ Amselem, M.\ Kleinmann, and M.\ \.Zukowski for their useful comments.




\begin{thebibliography}{00}
\bibitem{note-1}
Additional bounds are the maximum quantum violations of the Bell inequalities introduced in Ref. \onlinecite{BG03}, $2 \sqrt{N}$, with integer $N \ge 2$.
\end{thebibliography}
\end{document}



\title{Classical physics and the bounds of quantum correlations: \\ Supplemental Material}


\author{Diego Frustaglia}
\affiliation{Departamento de F\'{\i}sica Aplicada II, Universidad de Sevilla, E-41012 Sevilla, Spain}

\author{Jos\'e P. Baltan\'as}
\affiliation{Departamento de F\'{\i}sica Aplicada II, Universidad de Sevilla, E-41012 Sevilla, Spain}

\author{Mar\'{\i}a C. Vel\'azquez-Ahumada}
\affiliation{Departamento de Electr\'onica y Electromagnetismo, Universidad de Sevilla, E-41012 Sevilla, Spain}

\author{Armando Fern\'andez-Prieto}
\affiliation{Departamento de Electr\'onica y Electromagnetismo, Universidad de Sevilla, E-41012 Sevilla, Spain}

\author{Aintzane Lujambio}
\affiliation{Departamento de Electr\'onica y Electromagnetismo, Universidad de Sevilla, E-41012 Sevilla, Spain}

\author{Vicente Losada}
\affiliation{Departamento de F\'{\i}sica Aplicada I, Universidad de Sevilla, E-41011 Sevilla, Spain}

\author{Manuel J. Freire}
\affiliation{Departamento de Electr\'onica y Electromagnetismo, Universidad de Sevilla, E-41012 Sevilla, Spain}

\author{Ad\'an Cabello}
\affiliation{Departamento de F\'{\i}sica Aplicada II, Universidad de Sevilla, E-41012 Sevilla, Spain}




\maketitle



\subsection*{CLASSICAL MODELS FOR QUANTUM-LIKE PROBABILITIES}

Here we present two different models employing classical waves that allow an observer to operationally define discrete events (outcomes of sequential measurements of compatible observables) with probability distributions indistinguishable from the quantum ones. In particular, sharing the same exact bounds of correlations. These models are independent from the existence of physical quanta and the fact that a coarse-graining of quantum theory can result in a classical wave theory. 

\vspace{0.5cm}

{\bf Model 1:} Imagine the microwave experiment described in the manuscript, but with classical microwave detectors placed at each output terminal that click only after an energy threshold $E_0$ is surpassed. After a click, the detectors are reset back to zero. Different detectors would click at different instants as the microwave intensity is not equally distributed among the output terminals. Otherwise, the possibility of simultaneous clicks is removed by the introduction of a $\Delta E$ in the energy threshold.  
\vspace{0.5cm}

{\bf Model 2:} Imagine the microwave experiment with the intensity detectors described in the manuscript, but with a mechanism that loads an $N$-sided die according to the microwave intensity distribution over $N$ output ports. Each face of the die is identified with one output port. The die is made such that all the probabilities that a certain face is the uppermost when the die comes to rest after a throw are equal to the normalized microwave intensities at the outputs. Once the die is loaded we throw it to produce an outcome. 
\vspace{0.5cm}

%

\subsection*{EXPERIMENTAL SETUP}

The experimental setup consists in circuit tree networks of interconnected beam-splitters working in the microwave range (Fig. 2 and Supplemental Fig. S5). Each beam-splitter is implemented as a rat-race (hybrid ring) coupler [S1] designed to work at the center frequency of $2.45$ GHz, which corresponds to one of the ISM (Industrial, Scientific and Medical) radio bands defined by the International Telecommunication Conference. The rat-race coupler is essentially a 3dB microwave coupler with four matched ports that can be also used to obtain the sum or the difference of two signals. Thirty rat-race couplers were fabricated on a microwave substrate (ARLON AD 1000, $10.2$ dielectric constant, $1.27$ mm thickness) with a LPKF laser milling machine for PCB. SMA jack end launchers were soldered at the four ports of the couplers and 50-Ohms SMA terminations were screwed at unused ports. Some of the couplers are used as power dividers. These couplers only use three of the four ports, and the unused port must be matched with a 50 Ohms termination load to avoid reflections. The different couplers were interconnected with low-loss 50-Ohms coaxial cables 6" long (Pasternack, PE-SR405FL). Additionally, one single unequal split branch line coupler was fabricated with the same technology for the production of a particular microwave state (Supplemental Fig. S3). Measurements were done with an Agilent PNA series E8363B automatic vector network analyzer.

\subsection*{ERROR TREATMENT} 

We measure the transmission coefficients with three significant digits. The main error sources are the large number of rat-race couplers and coaxial cable segments with fluctuating electrical length, leading to microwave phase and power-loss fluctuations. The effect of bulk power loss is minimized by construction due to the symmetric design of the tree structures (Supplemental Fig. S6). Errors in the measured values of $E$ defined in (1), $M$ defined in (2), and $\chi$ defined in (3), are determined from imbalanced outcomes in the output channels. Moreover, ideal NC inequalities are corrected by introducing deviation rates $0 \le \xi \le 1$ in a series of additional tests (see below), defined as the fraction of faulty outcomes with respect to the ideal case where context-independence and order-independence of marginal probabilities, together with repeatability and minimal disturbance, are fully satisfied. 

\subsection*{TESTS OF CONTEXT INDEPENDENCE, ORDER INDEPENDENCE, REPEATABILITY, AND MINIMAL DISTURBANCE}

We tested nineteen representative sequences of observables relevant for evaluating $\chi$ (and $E$) in a tree arrangement over the eleven classical microwave states listed in Supplemental Table S1, organized in three groups:
\begin{itemize}
\item[(i)] $\alpha \beta \delta$, $ \alpha \delta \beta$, $ \delta \alpha \beta$, $ \delta \beta \alpha$, $ \beta \delta \alpha$, and $ \beta \alpha \delta$.
\item[(ii)] $AAA$, $BBB$, $DDD$, $aaa$, $bbb$, $ddd$, $\alpha \alpha \alpha$, $\beta \beta \beta$, and $ \delta \delta \delta$.
\item[(iii)] $\alpha A \alpha$, $\alpha a \alpha$, $\alpha \beta \alpha$, and $\alpha \delta \alpha$.
\end{itemize}
For group (i), we measured the sequential joint averages $\langle \dots \rangle$ and found deviations of the order of 1\% around the mean value 0.979. For groups (ii) and (iii), instead, we compared the results of the first measurement with those of the third one and evaluated up to which extent these results repeat. The largest deviation, with a value of 14\%, corresponds to the state $| \psi_{11} \rangle$ and the tree $bbb$. We then choose a global deviation rate $\xi_\chi=0.14$ coincident with the worst case.

Notice that the operators used to test $E$ are a subset of those used to test $\chi$. This subset includes
operator $b$, which produces the largest deviation rate obtained in
the state-independent NC tests described above corresponding to the
tree $bbb$. As a consequence, we choose $\xi_E=0.14$ as in
the previous, state-independent case.

We tested twelve representative sequences of observables relevant for evaluating $M$ in a tree arrangement over different sets of classical microwave states. The sequences are organized in three groups:
\begin{itemize}
\item[(iv)] $ABc$, $AcB$, $BAc$, $BcA$, $cAB$, and $cBA$.
\item[(v)] $AAA$ and $bbb$.
\item[(vi)] $ABA$, $ACA$, $AaA$, and $AcA$.
\end{itemize}
For group (iv), we measured the sequential joint averages $\langle \dots \rangle$ on a classical GHZ state and found deviations of the order of 0.5\% around the mean value 0.997. For groups (v) and (vi), instead, we compare the results of the first measurement with those of the third one and evaluate up to which extent these results repeat over four different classical states: GHZ, $|000\rangle$, $|111\rangle$ and $(|010\rangle -|001\rangle)/\sqrt{2}$. The largest deviation, with a value of 3\%, corresponds to the state $(|010\rangle -|001\rangle)/\sqrt{2}$ and the tree $bbb$. We then choose a global deviation rate $\xi_M=0.03$ coincident with the worst case.

\subsection*{VIOLATION ROBUSTNESS AGAINST IMPERFECTIONS}

Departures from ideal experimental conditions are accounted by deviation rates $\xi_E$, $\xi_M$ and $\xi_\chi$, leading to corrected versions of inequalities $E \le 2$, $M \le 2$, and $\chi \le 4$. The first one is rewritten as $E \le (1-\xi_E) \times 2 + \xi_E \times 4$, where $\xi_E$ is the corresponding deviation rate, $2$ is the upper bound of $E$ in an ideal experiment, and $4$ is the maximal algebraic value of $E$. This assumes the worst-case scenario, finding
\begin{equation}
E \le 2.28
\label{CHSH-corr}
\end{equation}
for noncontextual hidden-variable theories with $\xi_E=0.14$. Similarly, by defining $\chi \le (1-\xi_\chi) \times 4 + \xi_\chi \times 6$ with $\xi_\chi=0.14$ we find
\begin{equation}
\chi \le 4.28.
\label{PM-corr}
\end{equation}
Finally, for $M \le (1-\xi_M) \times 2 + \xi_M \times 4$ with $\xi_M=0.03$ we obtain
\begin{equation}
M \le 2.06.
\label{MB-corr}
\end{equation}


\begin{figure*}[!t]
\center
\includegraphics[scale=0.55]{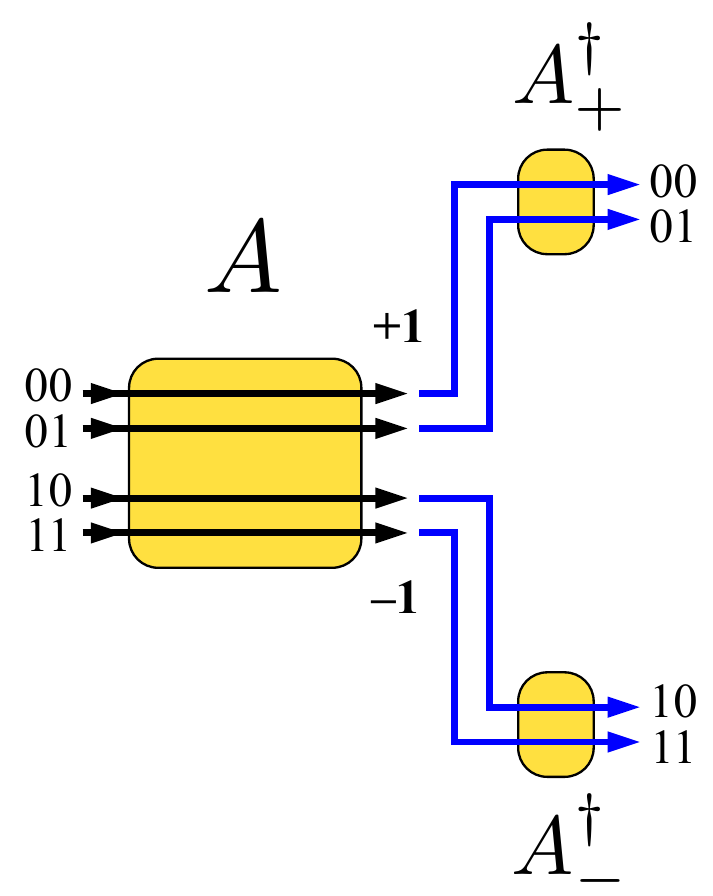} \hspace{2cm}
\includegraphics[scale=0.55]{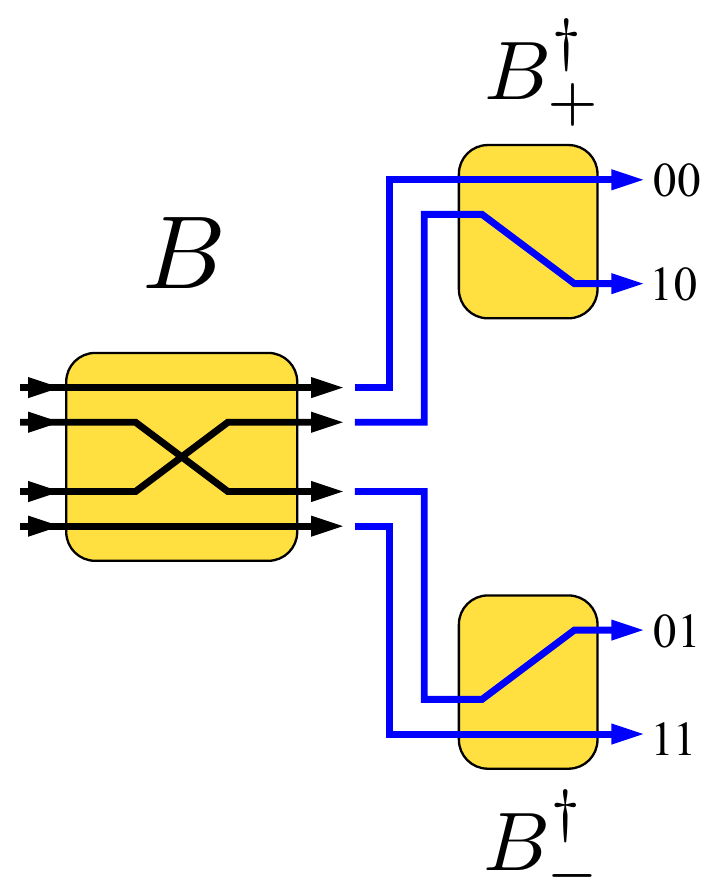} \hspace{2cm}
\includegraphics[scale=0.55]{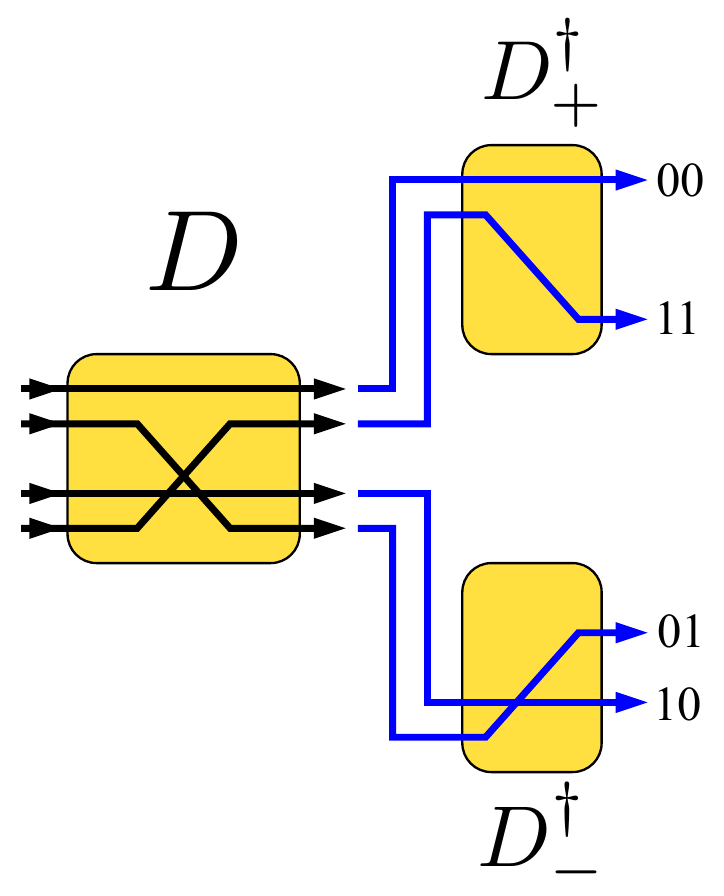} \\ \vspace{0.75cm}

\includegraphics[scale=0.55]{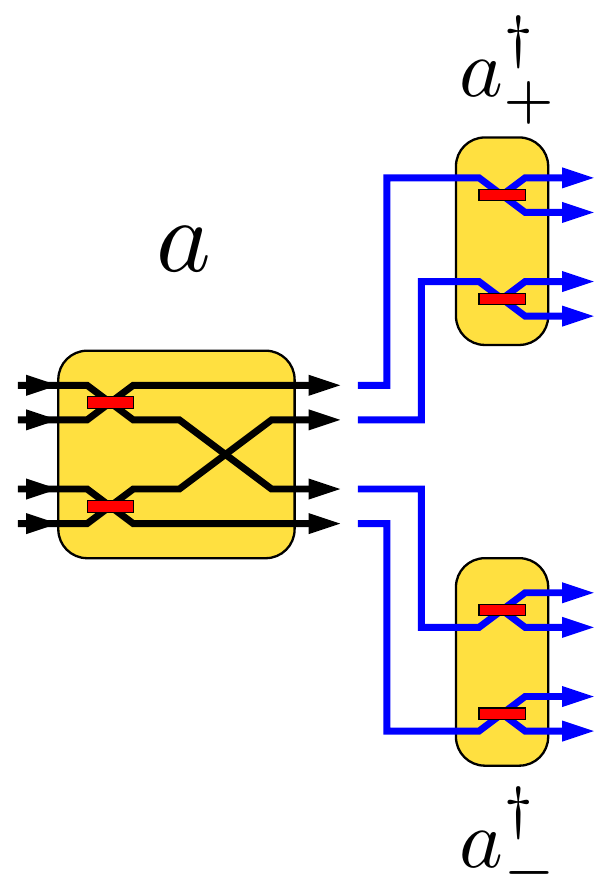} \hspace{2cm}
\includegraphics[scale=0.55]{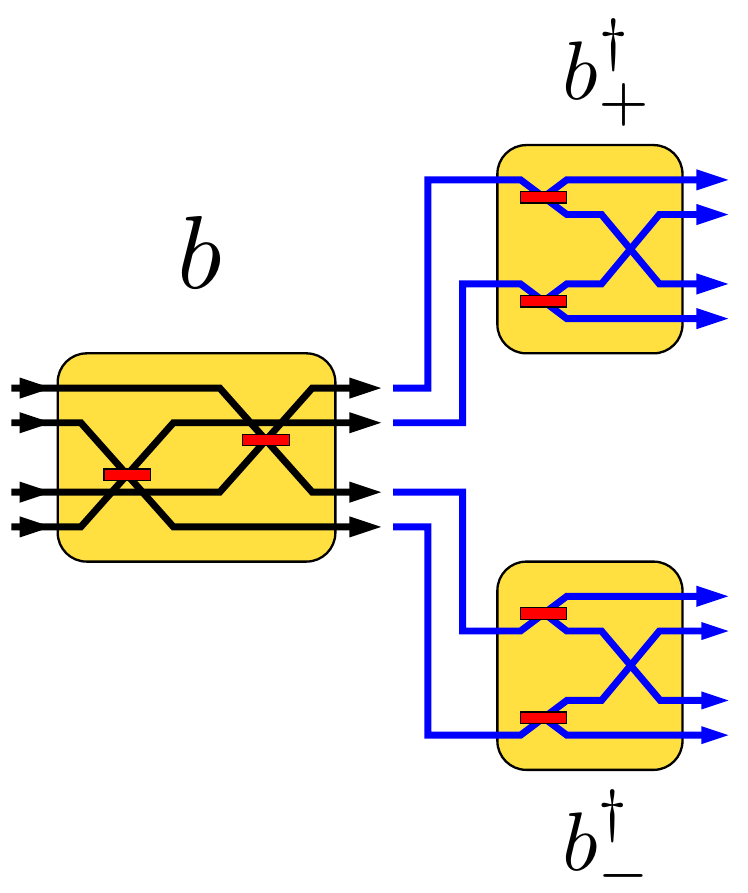} \hspace{2cm}
\includegraphics[scale=0.55]{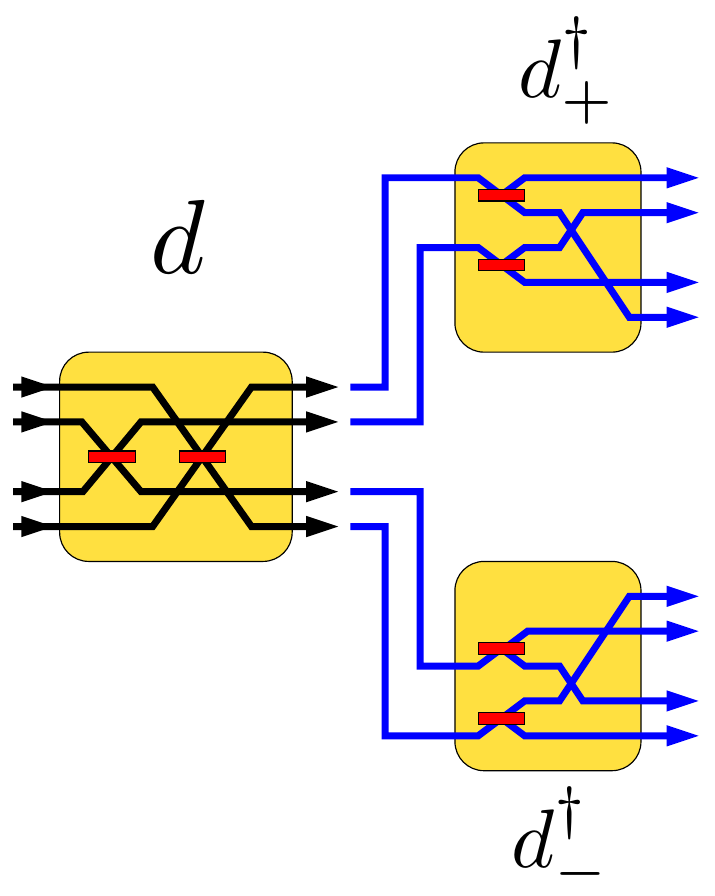} \\ \vspace{0.75cm}

\includegraphics[scale=0.55]{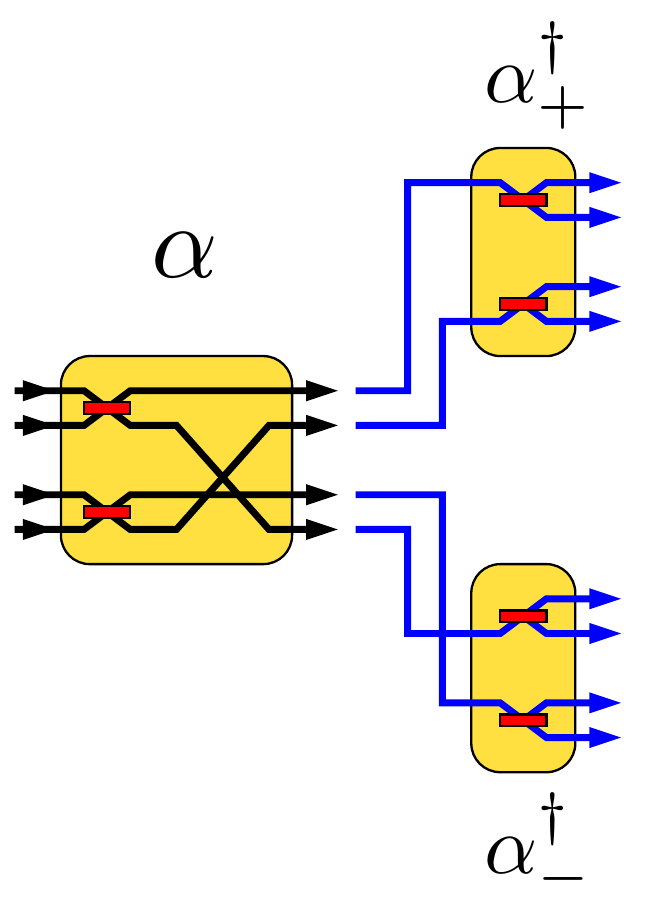} \hspace{1.5cm}
\includegraphics[scale=0.55]{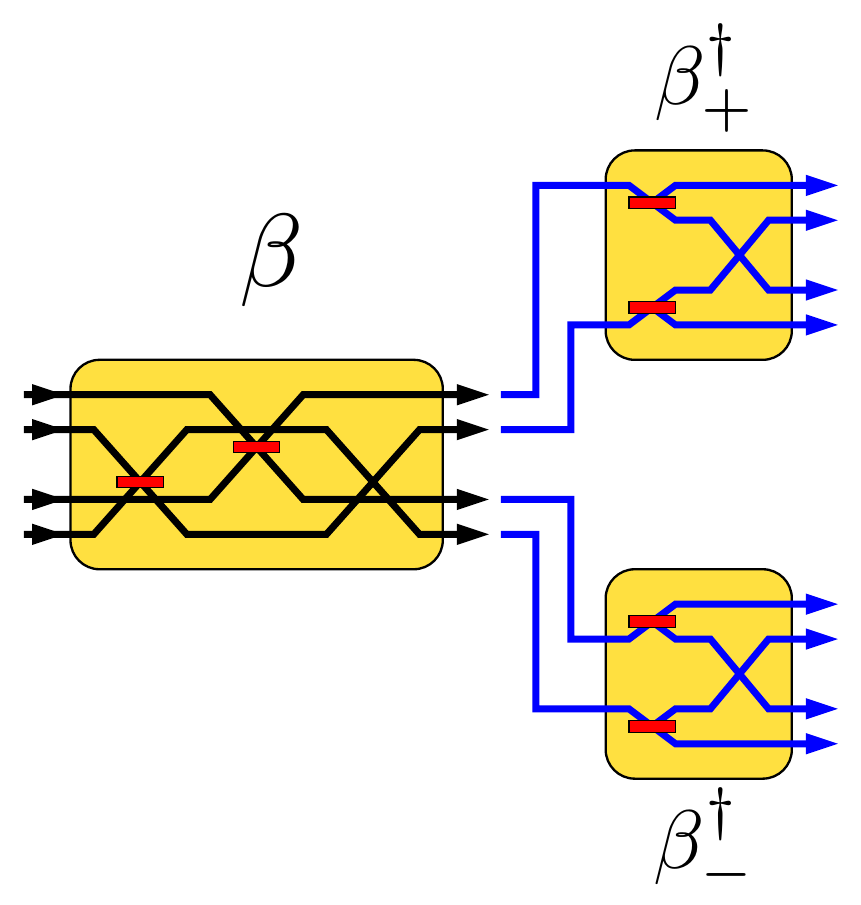} \hspace{1.5cm}
\includegraphics[scale=0.55]{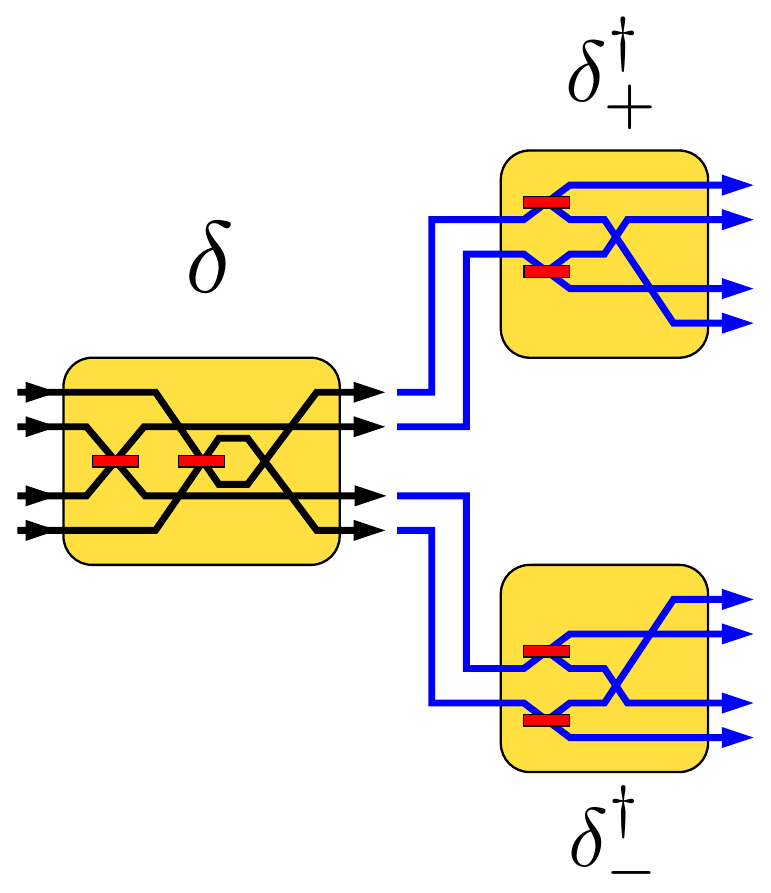}

\caption{Path representation of the observables $A=\sigma_z \otimes \mathbb{I}$, $B=\mathbb{I} \otimes \sigma_z$, $D=\sigma_z \otimes \sigma_z$, $a=\mathbb{I} \otimes \sigma_x$, $b=\sigma_x \otimes \mathbb{I}$, $d=\sigma_x \otimes \sigma_x$, $\alpha =\sigma_z \otimes \sigma_x$, $\beta =\sigma_x \otimes \sigma_z$, and $\delta=\sigma_y \otimes \sigma_y$ used to test $\chi$, including state recomposition. The subset $\{A,B,a,b\}$ is also used to test $E$. All path segments have equal length and the beam splitters work as indicated in Fig.~2 (b). Circuit~$A$ (upper left) includes detailed input path encoding. Upper (lower) output branches correspond to $+1$ ($-1$) outcomes.
}
\label{fig-S1}
\end{figure*}


\begin{figure*}[!t]
\center
\includegraphics[scale=0.5]{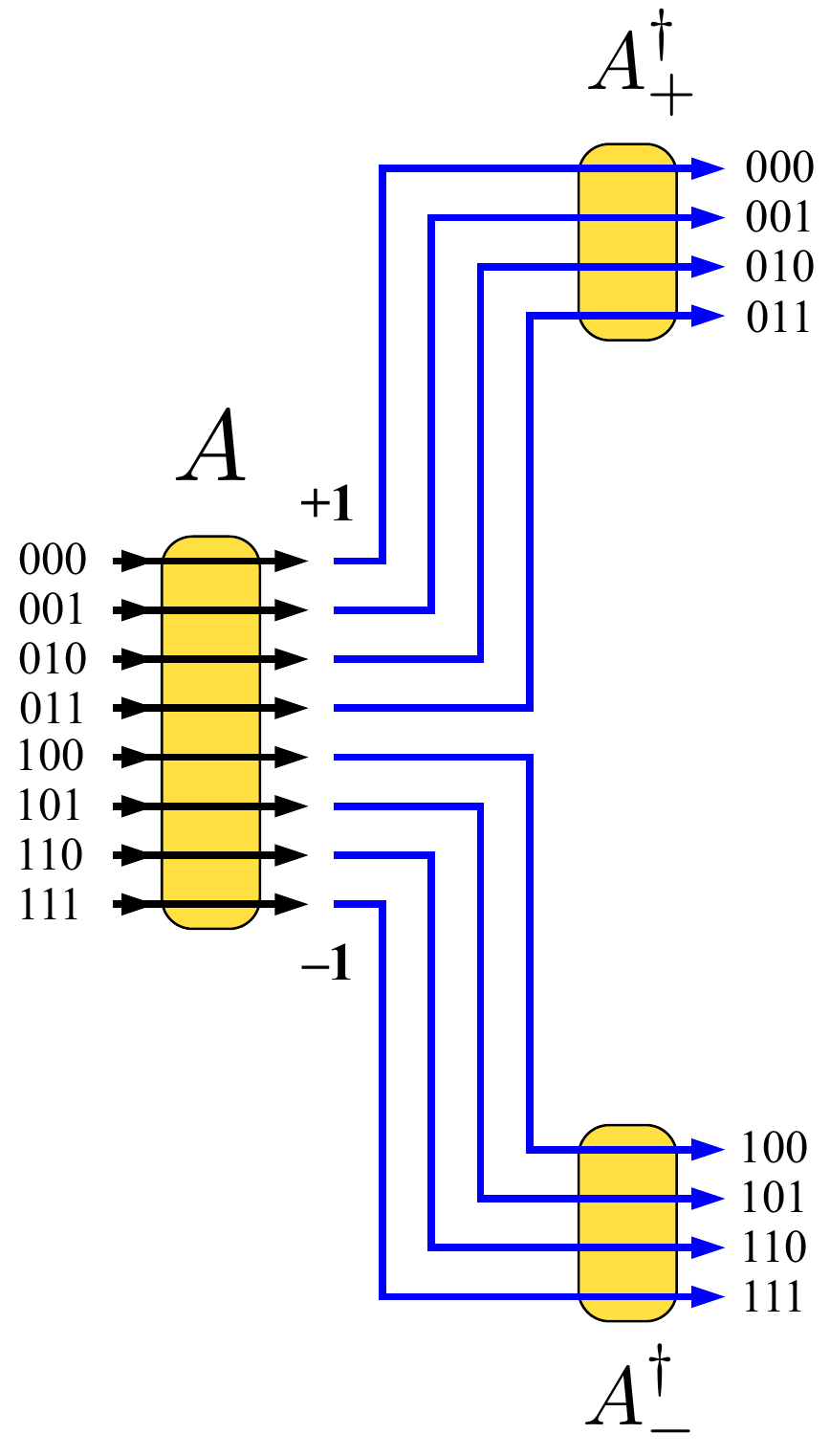} \hspace{1.5cm}
\includegraphics[scale=0.5]{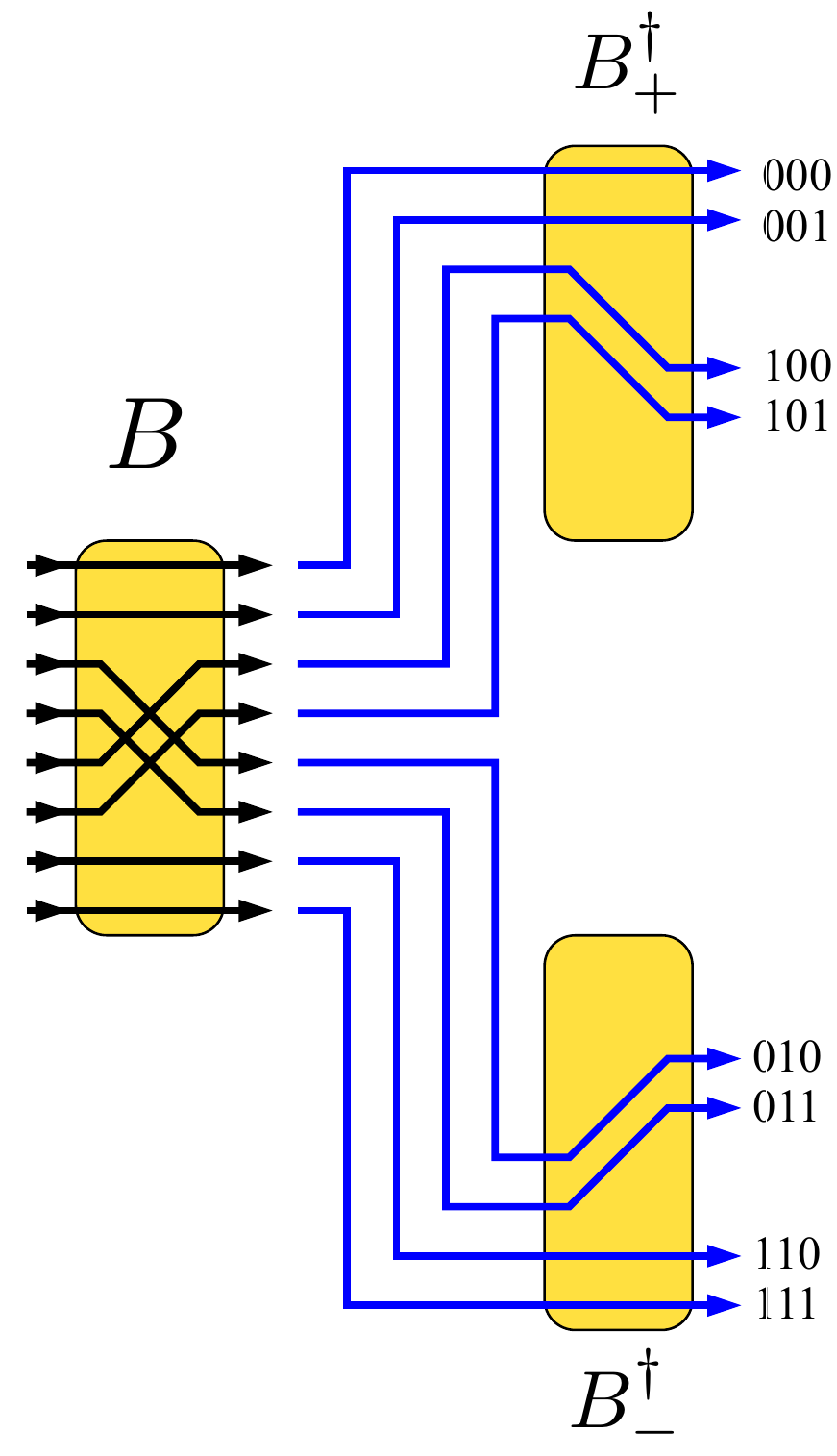} \hspace{1.5cm}
\includegraphics[scale=0.5]{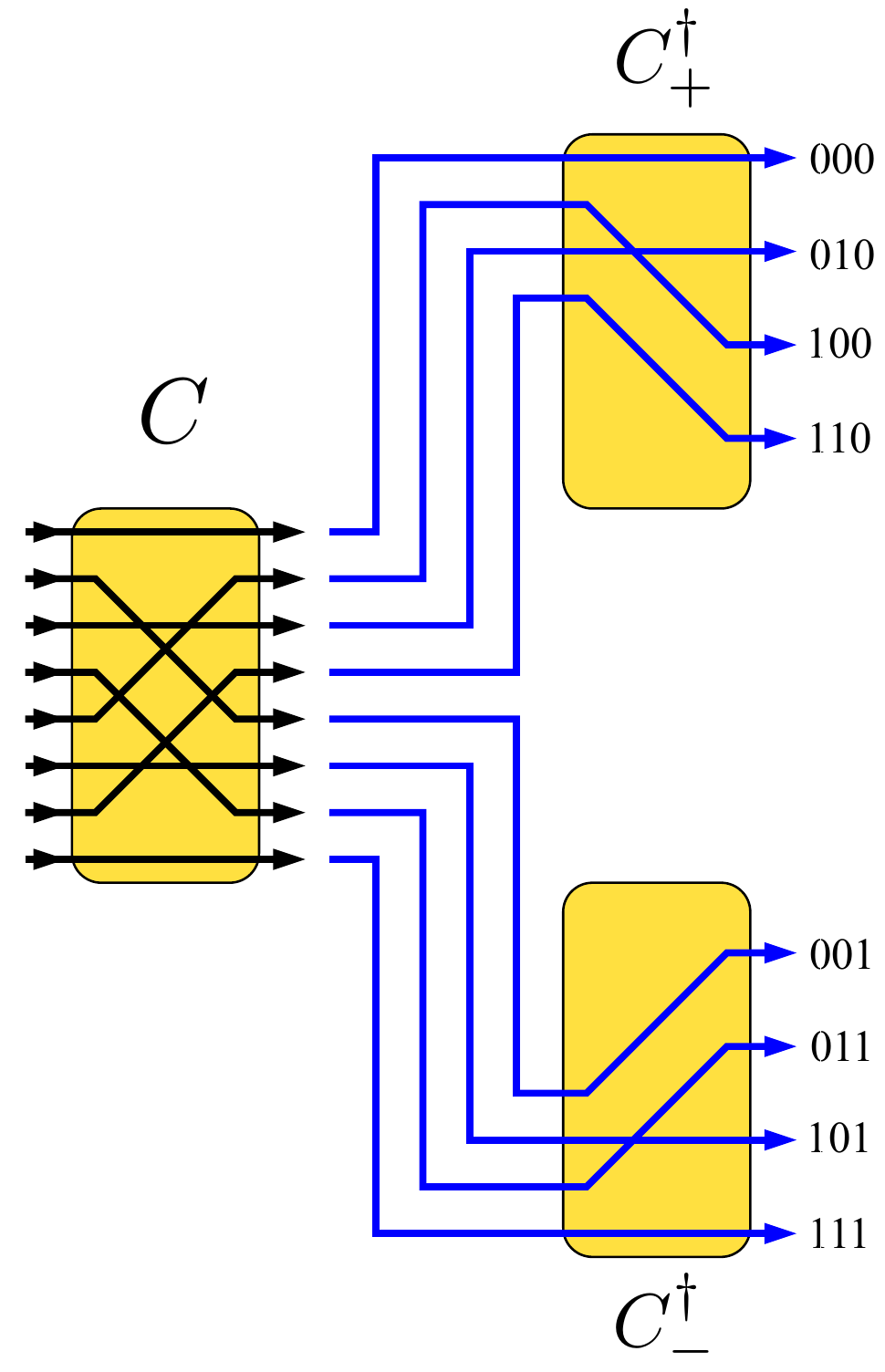} \\ \vspace{0.75cm}

\includegraphics[scale=0.5]{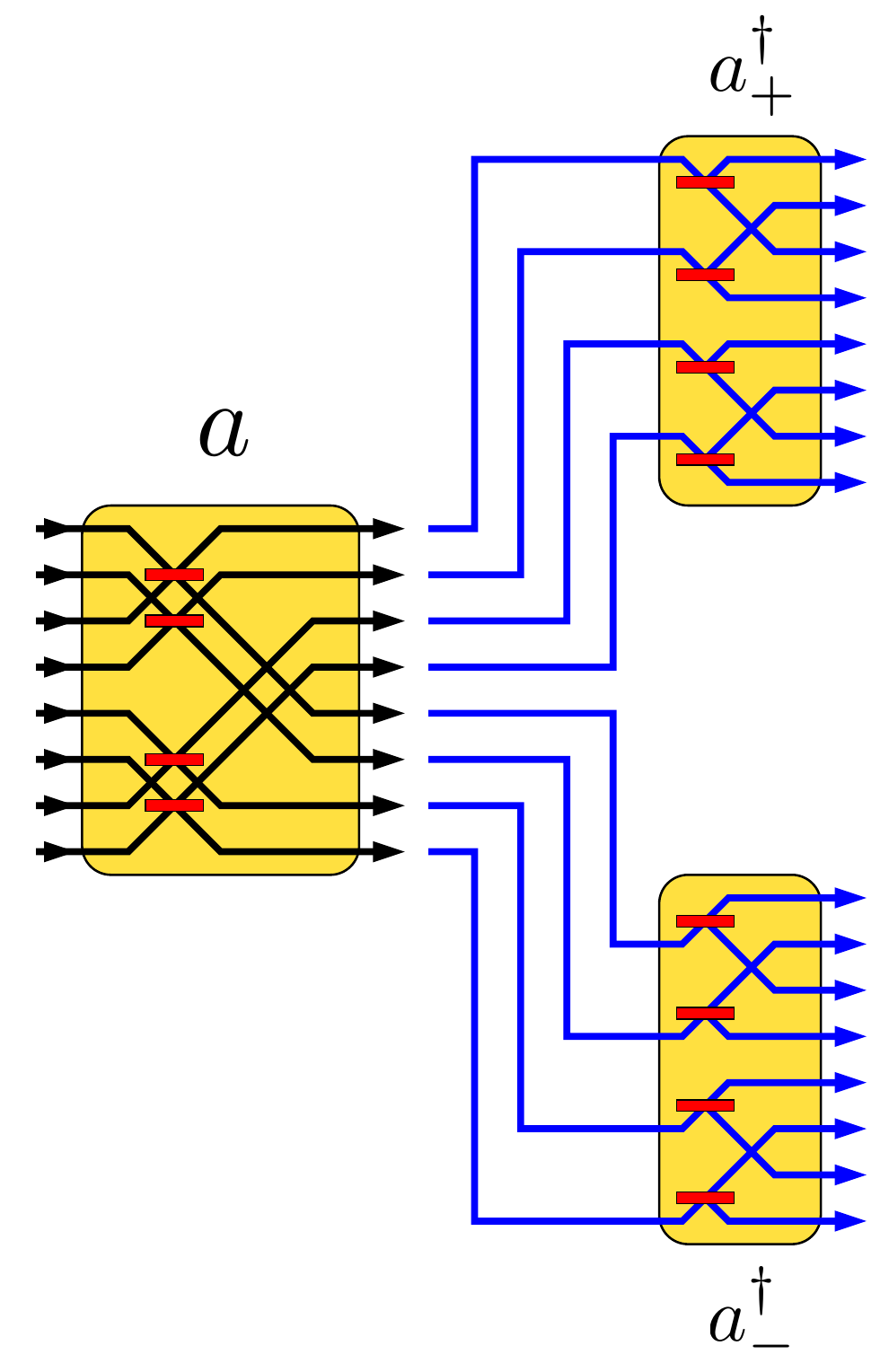} \hspace{1cm}
\includegraphics[scale=0.5]{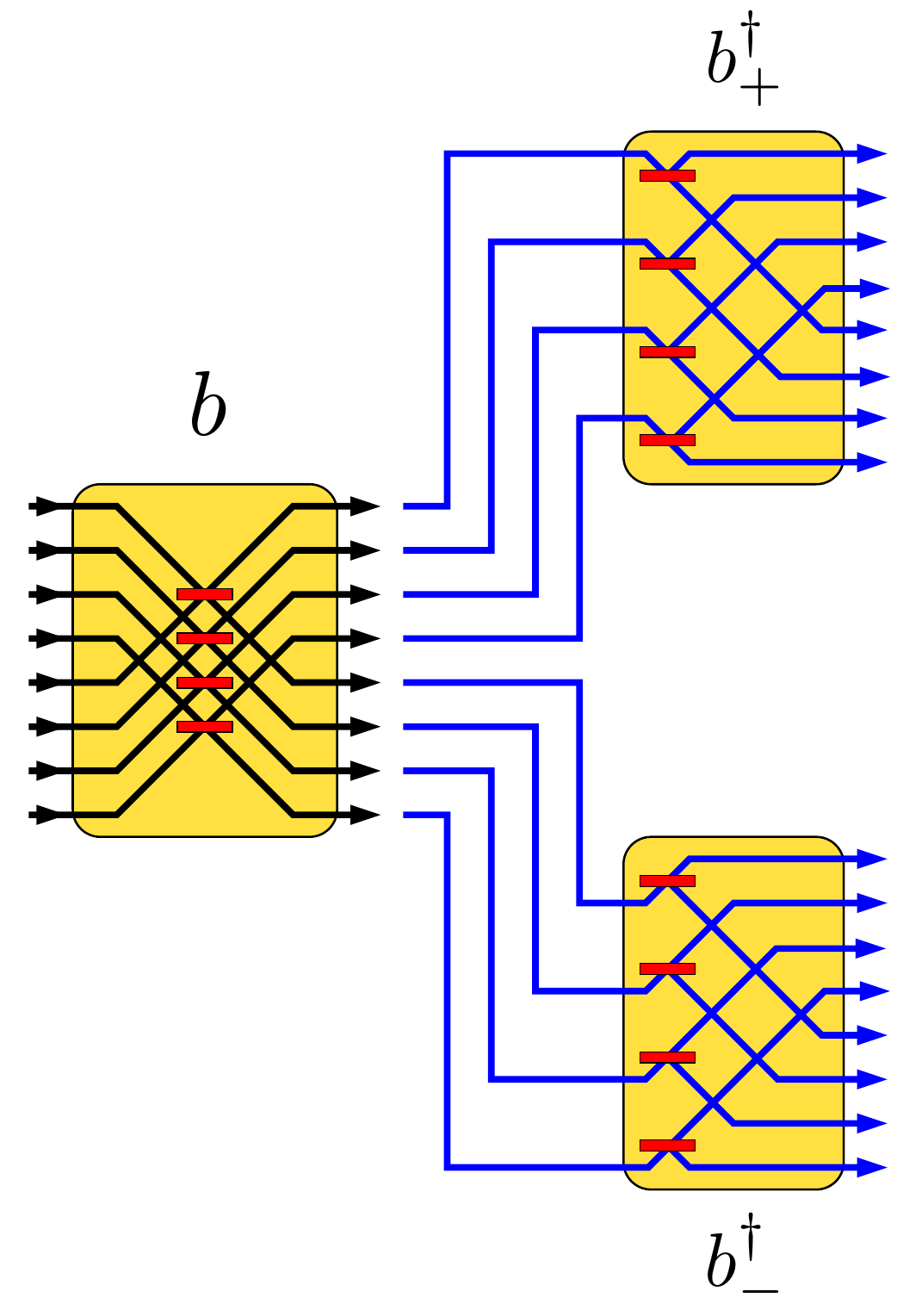} \hspace{1cm}
\includegraphics[scale=0.5]{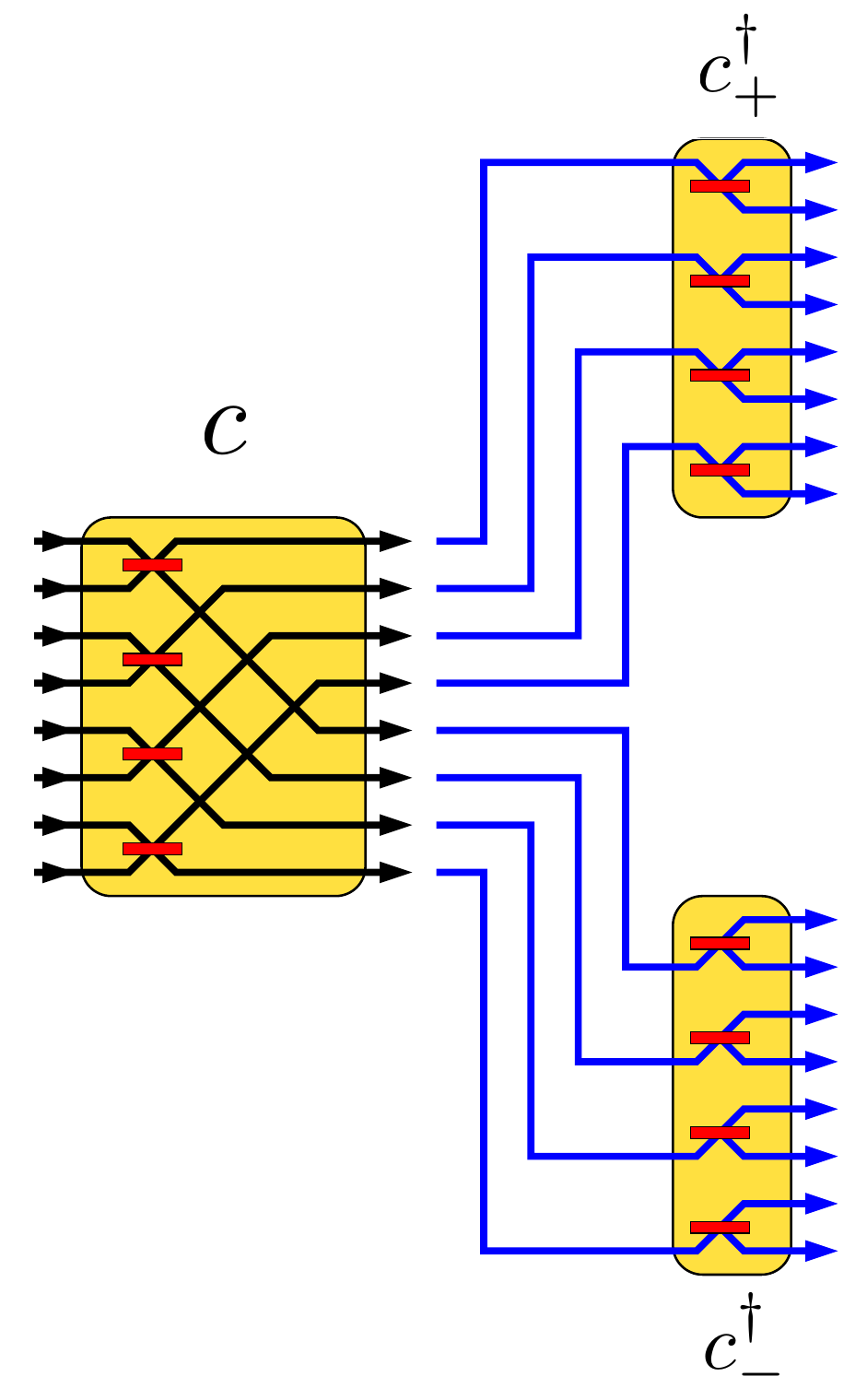}
\caption{Path representation of the observables $A=\sigma_z \otimes \mathbb{I} \otimes \mathbb{I}$, $B=\mathbb{I} \otimes \sigma_z \otimes \mathbb{I}$, $C=\mathbb{I} \otimes \mathbb{I} \otimes \sigma_z$,
 $a=\mathbb{I} \otimes \sigma_x \otimes \mathbb{I}$, $b=\sigma_x \otimes \mathbb{I} \otimes \mathbb{I}$, and $c=\mathbb{I} \otimes \mathbb{I} \otimes \sigma_x$ used to test $M$, including state recomposition, in $2\times2\times2$ dimensions. All path segments have equal length and the beam splitters work as indicated in Fig.~2 (b). Circuit~$A$ (upper left) includes detailed input path encoding. Upper (lower) output branches correspond to $+1$ ($-1$) outcomes.
}
\label{fig-S2}
\end{figure*}


\begin{figure*}[!t]
\center
\includegraphics[scale=0.7]{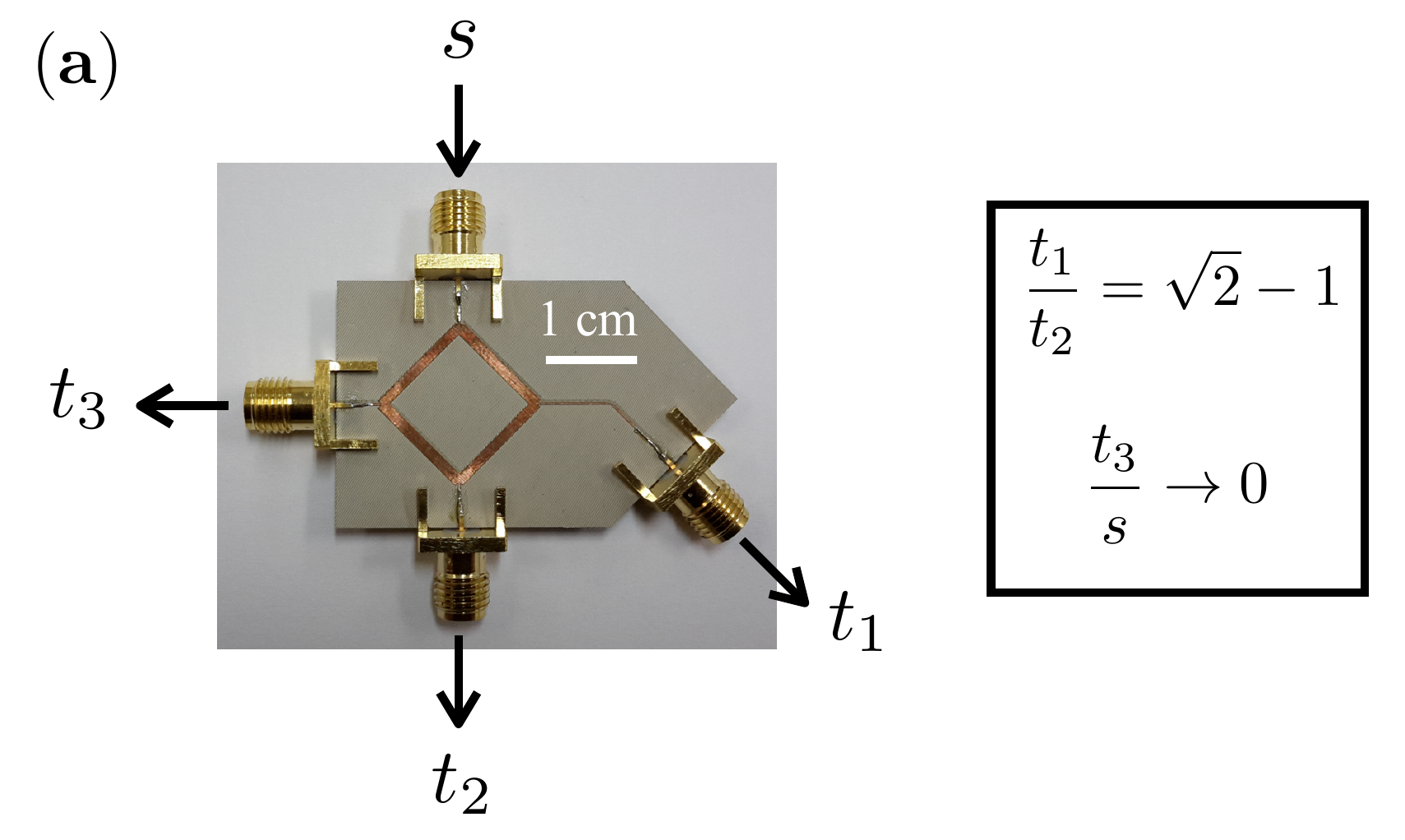} \\ \vspace{0.75cm}
\includegraphics[scale=0.6]{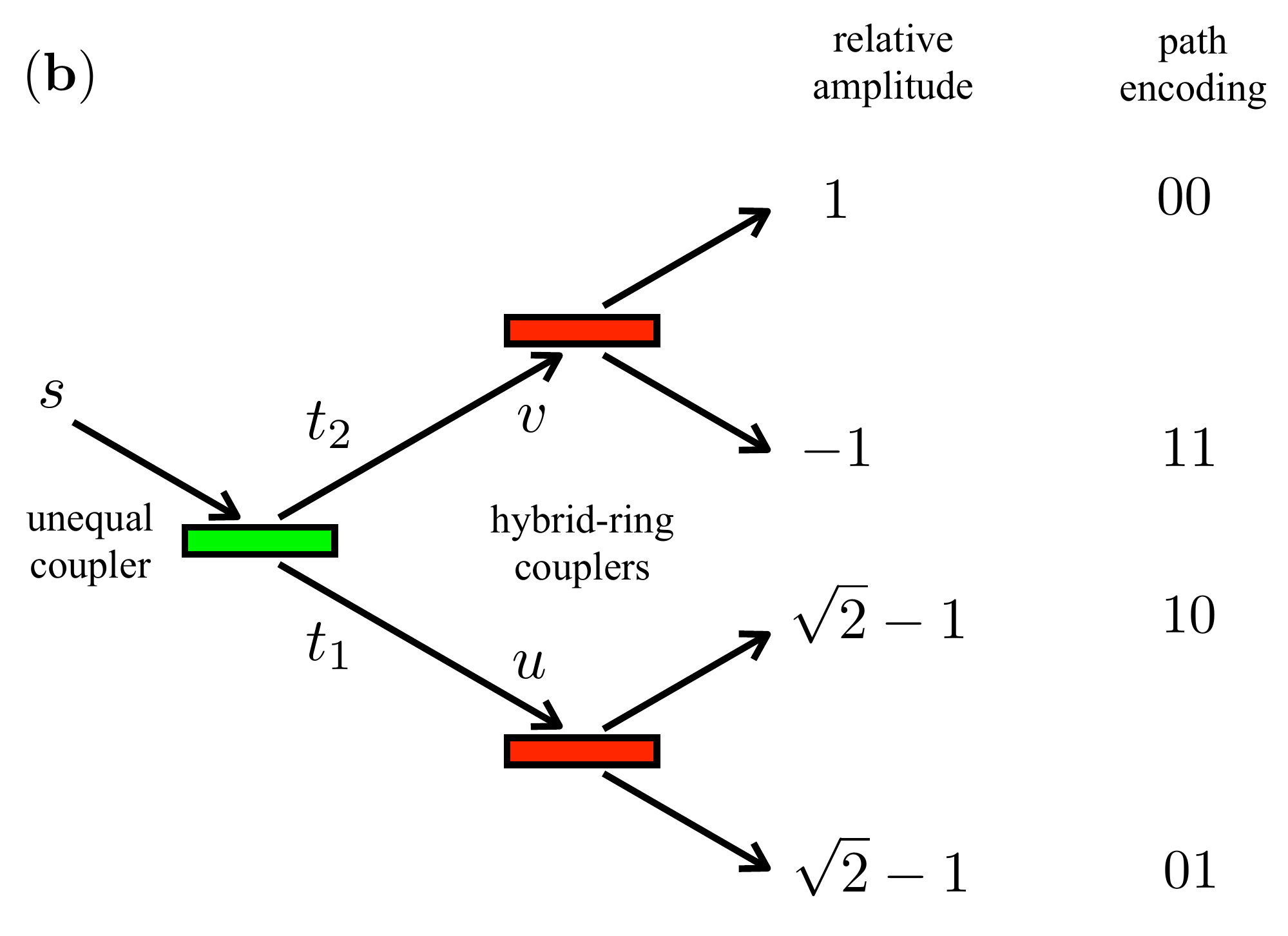}
\caption{ (a) Unequal split branch line coupler designed for working at $2.45$ GHz, acting as a classical microwave beam-splitter for producing the state $|\Psi_{\rm CHSH} \rangle = \left[|00\rangle-|11\rangle+(\sqrt{2}-1)(|01\rangle+|10\rangle)\right]/(2\sqrt{2-\sqrt{2}})$. The incoming signal $s$ is unequally split into two main outcome signals $t_1$ and $t_2$. A residual signal $t_3$ is derived to an isolated port. (b) Scheme representing the production of state $|\Psi_{\rm CHSH} \rangle$ from an incoming signal $s$ by multiple splitting through one unequal coupler [panel (a)] and two hybrid-ring couplers [Fig. 2 (b) and (c)]. 
}
\label{fig-S3}
\end{figure*}


\begin{figure*}[!t]
\center
\includegraphics[scale=0.7]{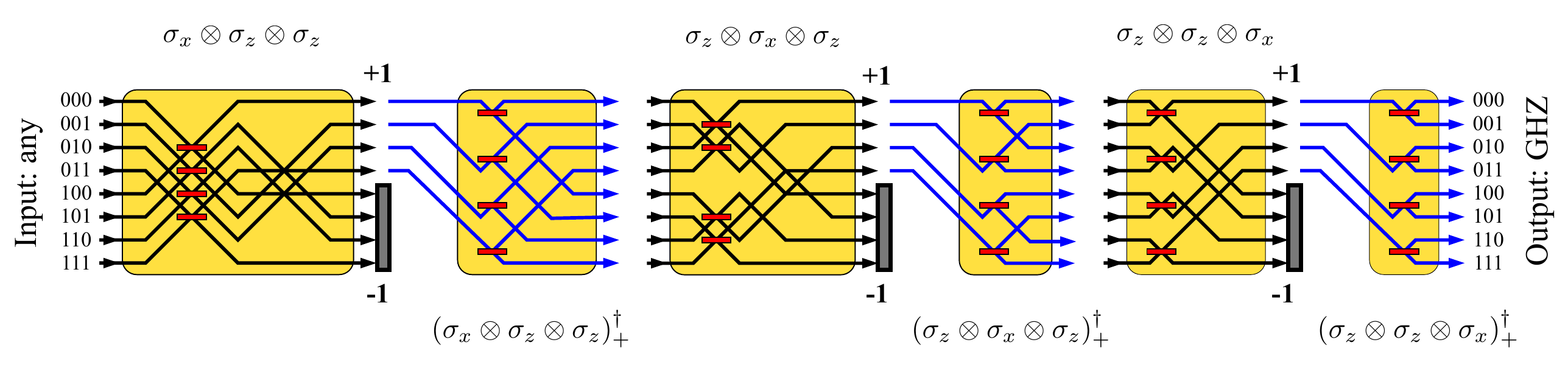}
\caption{Classical GHZ-like state preparation. For any input signal, a finite output gives $|\Psi_{\rm GHZ} \rangle= (|000\rangle +|001\rangle+|010\rangle-|011\rangle +|100\rangle-|101\rangle-|110\rangle -|111\rangle)/\sqrt{6}$ up to a global phase and a normalization factor. All path segments have equal length and the beam splitters work as indicated in Fig.~2 (b).
}
\label{fig-S4}
\end{figure*}


\begin{figure*}[!t]
\center
\includegraphics[scale=0.7]{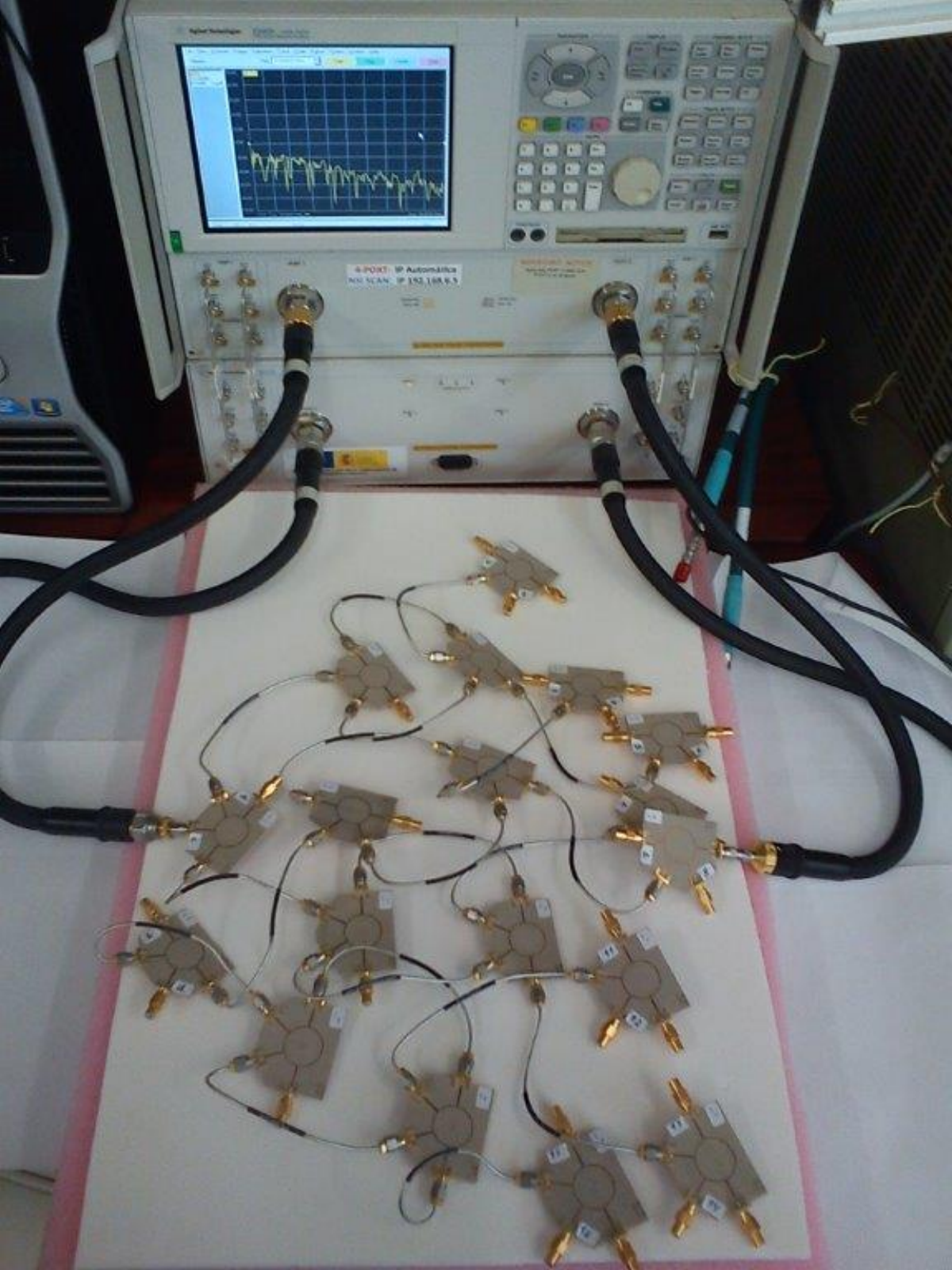}
\caption{Classical-microwave circuit tree network built upon two basic elements: coaxial-cable segments of equal electrical length and hybrid-ring (rat-race) couplers used as beam-splitters designed to work at $2.45$ GHz, as detailed in Fig.~2 (c). Microwave transmission coefficients between left and right terminals are measured by an automatic vector network analyzer.
}
\label{fig-S5}
\end{figure*}


\begin{figure*}[!t]
\center
\includegraphics[scale=0.7]{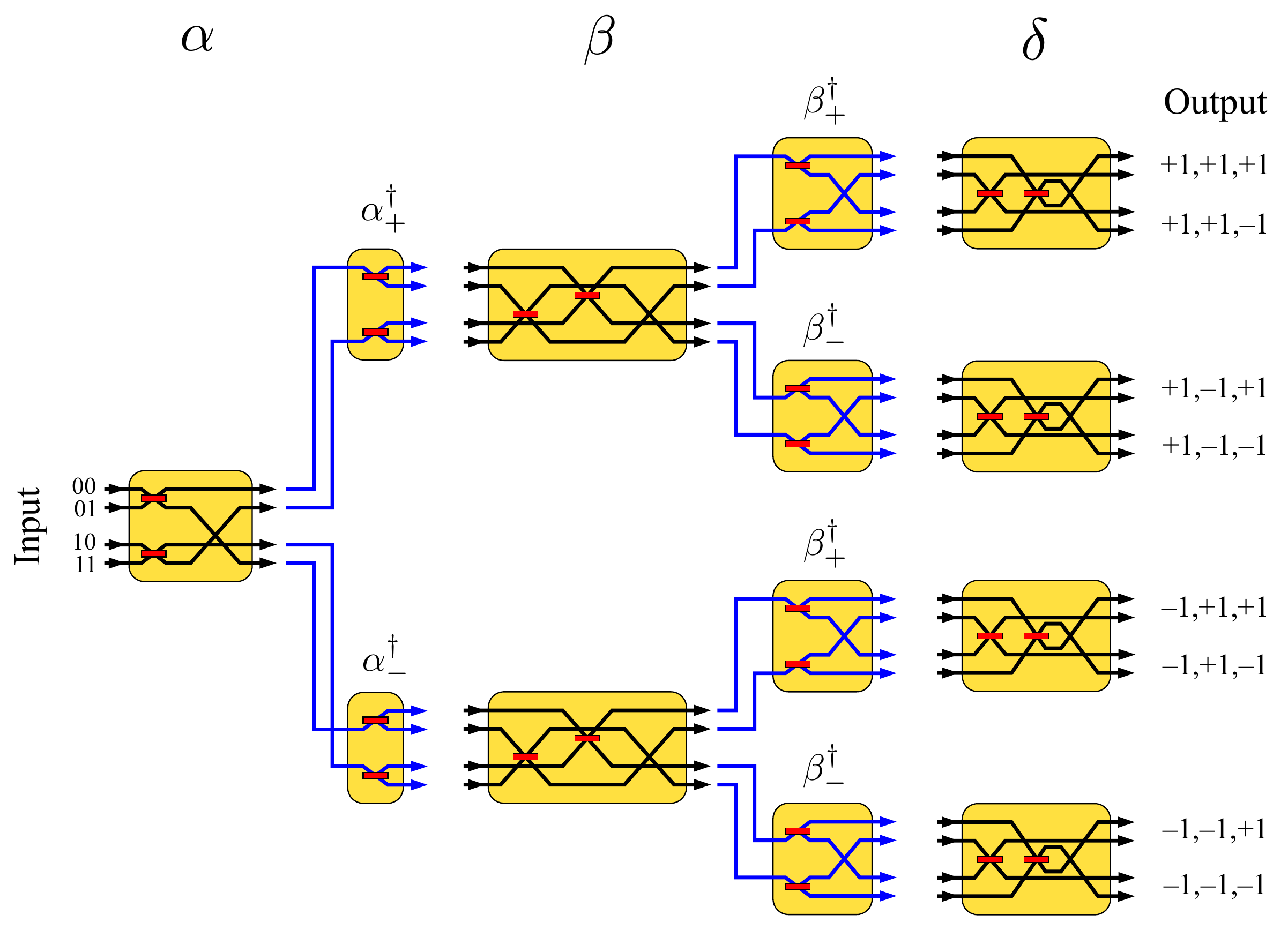}
\caption{Schematic representation of the circuit tree network used to measure $\langle \alpha \beta \delta \rangle$ for testing $\chi$, Eq.(3). All path segments have equal length and the beam splitters work as indicated in Fig.~2 (b). Schemes corresponding to any other correlator are arranged in a similar manner.}
\label{fig-S6}
\end{figure*}


\begin{table}[htdp]
\caption{Experimental values of $\chi$ for eleven different classical microwave states.}
\begin{center}
\begin{tabular}{|l|c|}
\hline
~~~~~~~~~~~~~~~~~~~~Input state & $\chi$ \\
\hline
$|\psi_1 \rangle = |00 \rangle$ & $5.93(14)$\\
$|\psi_2 \rangle = |01 \rangle$ & $5.93(17)$ \\
$|\psi_3 \rangle = |10 \rangle$ & $5.94(14)$ \\
$|\psi_4 \rangle = |11 \rangle$ & $5.93(21)$ \\
$|\psi_5 \rangle = \frac{1}{\sqrt{2}}(|00 \rangle + |11 \rangle)$ & $5.93(21)$ \\
$|\psi_6 \rangle = \frac{1}{\sqrt{2}}(|00 \rangle - |11 \rangle)$ & $5.93(22)$ \\
$|\psi_7 \rangle = \frac{1}{\sqrt{2}}(|01 \rangle + |10 \rangle)$ & $5.94(20)$ \\
$|\psi_8 \rangle = \frac{1}{\sqrt{2}}(|01 \rangle - |10 \rangle)$ & $5.91(24)$ \\
$|\psi_9 \rangle = \frac{1}{2}(|00 \rangle +|01 \rangle + |10 \rangle+ |11 \rangle)$ & $5.93(22)$ \\
$|\psi_{10} \rangle = \frac{1}{2}(|00 \rangle -|01 \rangle + |10 \rangle+ |11 \rangle)$ & $5.90(44)$ \\
$|\psi_{11} \rangle =0.83~ |00 \rangle + 0.56~ e^{i 0.52 \pi}|11 \rangle$ & $5.93(44)$ \\
\hline
\end{tabular}
\end{center}
\label{table}
\end{table}


\subsection*{SUPPLEMENTAL REFERENCES}

[S1] Pozar, D. M.
 {\em Microwave Engineering}
 (Addison Wesley, New York, 1993).